\documentclass[twocolumn,amssymb, nobibnotes, showpacs, superscriptaddress, aps, prd]{revtex4-1}
\usepackage{amsmath}
\usepackage{epsfig}
\begin{document}
\title{All-Optical, High-Fidelity Polarization Gate Using Room-Temperature Atomic Vapor}


\author{Runbing Li}
\affiliation{National Institute of Standards and Technology, Gaithersburg, Maryland USA 20899}
\affiliation{State Key Laboratory of Magnetic Resonance and Atomic and Molecular Physics,
Wuhan Institute of Physics and Mathematics, Chinese Academy of Sciences, Wuhan 430071, China}
\affiliation{Center for Cold Atom Physics, Chinese Academy of Sciences, Wuhan 430071, China}
\author{Chengjie Zhu}
\affiliation{National Institute of Standards and Technology, Gaithersburg, Maryland USA 20899}
\affiliation{State Key Laborotary of Precision Spectroscopy and Department of Physics, East China Normal University, Shanghai 200062 China}
\author{L. Deng}
\affiliation{National Institute of Standards and Technology, Gaithersburg, Maryland USA 20899}
\author{E.W. Hagley}
\affiliation{National Institute of Standards and Technology, Gaithersburg, Maryland USA 20899}

\begin{abstract}
An all-optical atomic Controlled-NOT (CNOT)/polarization gate operation is demonstrated with low light intensities in a room-temperature atomic medium. Using a Polarization-Selective-Kerr-Phase-Shift (PSKPS) technique, a $\pi$ phase shift is written to only one of the two circularly-polarized components of a linearly-polarized input signal field by a weak phase-control field with "magic" detuning. At the exit of the medium, the signal field maintains its original strength but acquires a 90$^{\rm o}$ linear polarization rotation, demonstrating the first fast, high-fidelity CNOT/polarization gate operation in a room-temperature atomic medium. This development opens the realm of possibilities for potential future extremely low light level telecommunication and information processing systems.
\end{abstract}

\maketitle
\vskip 10pt
\noindent Efficient optical-field manipulation protocols at very low light intensities and low power consumption, are critically important to next-generation advanced telecommunications and information processing \cite{1,2,3,4}. To this end, optical Controlled-NOT (CNOT) gates and polarization gates using polarization-encoded optical fields are important vector-gate operations \cite{5,6,7,8,9,10,11,12,13,14,15,16,17} with which advanced telecommunication and quantum information processing systems can be built. Although various weak-light and extremely low power consumption schemes have already been proposed for optical logic operations \cite{14,15,16,17}, the demonstration of a fast, all-optical CNOT/polarization gate using atomic media at very weak gate-switching light levels has proved to be very challenging, even in the classical field limit. To date, no such weak light CNOT/ polarization gate has been experimentally realized using room-temperature atomic medium. The experimental results demonstrated in this work open the door for the first time to potential applications and implementation of extremely low control light optical logic operations \cite{18} in advanced telecommunication protocols and networks.

\vskip 10pt
\noindent One approach to realize weak-light optical logic gates and controlled gate operations is to build a universal vector gate, such as a CNOT gate or a polarization gate, which would function as the basic building block of next-generation telecommunication systems.  Recently, it has been shown that in a Raman-gain-based $N-$scheme \cite{19} a phase-control field can induce a sufficiently large signal field nonlinear Kerr phase shift. A phase gate based on the Kerr effect is essentially a scalar gate since the phase shift depends on the intensity rather than on the vector properties of the light field. The challenge is to construct a true vector gate operation using such a scalar gate so that the desired optical logic operations can be realized with very low gate-switching/control field intensities.

\vskip 10pt
\noindent In this paper, we demonstrate a novel polarization-selective-Kerr-phase-shift (PSKPS) protocol and an optical, high-fidelity complete atomic linear polarization gating operation in which the polarization state of a very weak signal field can be rapidly and orthogonally changed by a phase-control field with a "magic" detuning. The intensity of the polarization-switching field used in this high-fidelity CNOT/polarization gate operation is 2 mW/cm$^2$, which is equivalent to 0.5 nW, 3 $\mu$s optical pulse of wavelength $\lambda$ = 800 nm confined in a typical commercial photonic hollow-core fiber

\begin{figure}
  \centering
  \includegraphics[angle=90,width=3.25 in]{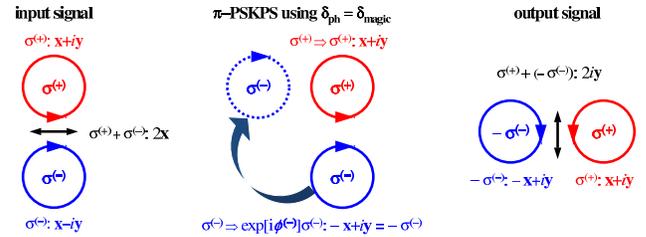}
  \caption{Fundamental principle of achieving a photonic polarization gate using a PSKPS technique. Left: Two circularly polarized components of a linearly polarized input field.  Middle: Selective Kerr phase shift of $\pi$ to the $\sigma^{(-)}$ component.  Right: Resultant polarization of the combined output field.}
\end{figure}

\vskip 10pt
\noindent Conceptually, the PSKPS protocol and the all-optical polarization gate operation demonstrated here are described in Fig.1. Here, the two circularly polarized components of a very weak linearly polarized input signal field are separately addressed by a weak phase-control light field, resulting in different nonlinear Kerr phase shifts $\phi^{(\pm)}$ being written to the different polarization modes $\sigma^{(\pm)}$.  If the differential nonlinear gain $\Delta G=G^{(+)}-G^{(-)}$ can be made negligibly small while simultaneously achieving $\phi^{(+)}$ = 0 and $\phi^{(-)}=\pi$, then at the exit of the medium the linear polarization of the signal field will have undergone a perfect 90$^o$ rotation. With this perfect polarization gate operation, a CNOT gate and a host of other fast, all-optical logic gates \cite{18} can be easily constructed.
\vskip 10pt
\noindent Theoretically, the polarization gate operation and the PSKPS technique can be best understood using a $^{87}$Rb atomic system shown in Fig. 2a.  We assume that initially all atoms are in the $|5S_{1/2}, F=1, m_{F}=+1\rangle$ state and a pump field (Rabi frequency $\Omega_P$) drives the $|5S_{1/2}, F=1, m_{F}=+1\rangle\rightarrow|5P_{3/2}, F^{'}=2, m_{F^{'}}=0\rangle$ transition with a large one-photon detuning $\delta$ to suppress spontaneous emission.  A very weak linear polarization-encoded signal field (Rabi frequency $\Omega_S$) couples the $|5P_{3/2}, F^{'}=2, m_{F^{'}}=0\rangle\rightarrow|5S_{1/2}, F=2\rangle$ transition.
A key consideration in this scheme is that atomic transition selection rules dictate that the direct $|F^{'}=2, m_{F^{'}}=0\rangle\rightarrow|F=2, m_F=0\rangle$ transition is forbidden when driven by the linearly-polarized signal field.  Consequently, the signal light can only couple, via its two circularly polarized components ($\Omega_{S\pm}$, red and blue arrows), i.e., the $|5P_{3/2}, F^{'}=2, m_{F^{'}}=0\rangle\rightarrow|5S_{1/2}, F=2, m_{F}=\pm 1\rangle$ transitions. The strategy for achieving a CNOT/polarization gate is thus to selectively introduce a nonlinear Kerr phase shift via a very weak phase-control field to only one signal coupling branch.  This selective phase shift allows the construction of a vector gate using a Kerr phase gate, and is the fundamental principle behind our CNOT/polarization gate operation.
\vskip 10pt
\noindent When a phase-control light field (Rabi frequency $\Omega_{ph}$ coupling the $|5S_{1/2}, F=2, m_{F}=\pm 1\rangle\rightarrow|5P_{1/2}, F^{''}=1,2\rangle$ transitions) is applied, both circularly polarized components of the signal field acquire a nonlinear gain $G^{(\pm)}$ and a phase shift $\phi^{(\pm)}$. Since initially the amplitudes of the circularly polarized components are exactly the same, simultaneously satisfying the following conditions ($N,M$ are integers),
\begin{eqnarray}
G^{(+)}=G^{(-)},\;\phi^{(+)}=2N\,\pi,\;{\,\rm and\;}\,\phi^{(-)}=(2M+1)\,\pi,
\end{eqnarray}
results in a perfect 90$^{\rm o}$ rotation of the signal polarization required for CNOT or polarization-gate operation.

\begin{figure}
  \centering
  \includegraphics[angle=90,width=3.25 in]{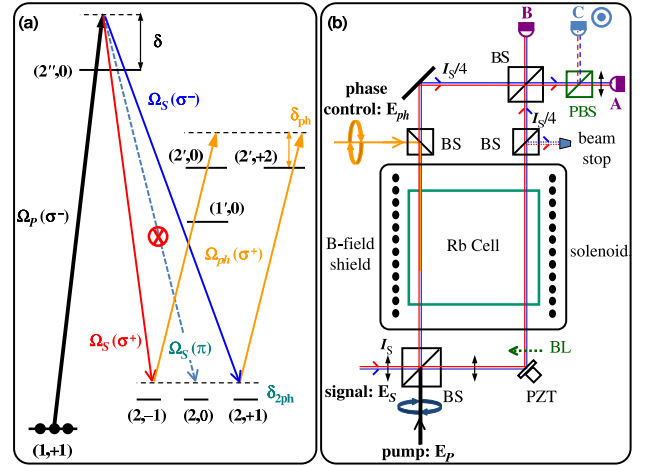}
  \caption{(a) Energy diagram and laser couplings of a lifetime broadened atomic system for the PSKPS-based photonic polarization gate operation. $\delta$: one-photon, pump-field detuning. $\delta_{2ph}$: two-photon detuning. $\delta_{\rm ph}$:  phase-control field detuning. Two downward arrows represent the two circular polarization components ($\Omega_{S\pm}$) of the linearly-polarized signal field ($\Omega_S$). The magic detuning $\delta_{\rm ph}=\delta_{\rm magic}$ is defined as the detuning that results in a complete orthogonal polarization rotation of the signal field with minimal phase-control ($\Omega_{ph}$) and pump ($\Omega_P$) fields. (b) Experimental setup.  BS: 50-50 beam splitter.  PBS: polarization beam splitter. BL: beam block. $A$, $B$, and $C$: detectors. PZT: piezo-controlled mirror.}
\end{figure}

\vskip 10pt
\noindent Physically, satisfying Eq. (1) requires that the phase-control light field interacts with different upper states for the different circular components at a specific phase-control light detuning.  We define a ``magic" detuning $\delta_{ph}=\delta_{\rm magic}$ as the phase-control field detuning at which a full and complete signal-field polarization rotation is achieved with minimal phase-control field intensity $I_{ph}$.  Figures 3a and 3b display the results of numerical calculations that show the existence of $\delta_{\rm magic}$ for the lifetime-broadened atomic system given in Fig. 2a. In Fig. 3a we show the differential gain $\Delta G=G^{(+)}-G^{(-)}$ and nonlinear Kerr phase shifts $\phi^{(\pm)}$ for the $\sigma^{(\pm)}$ modes of the signal field as a function of the phase-control light detuning.  Here, $\delta_{ph}=\delta_{\rm magic}$ indicates a special choice of the phase-control light detuning at which all three relations given in Eq. (1) are simultaneously satisfied. In Fig. 3b the magnitudes of $\Delta G$ and $\phi^{(\pm)}$ of the signal field are plotted in colored contour plots as functions of the pump-field detuning $\delta$ and the phase-control field detuning $\delta_{ph}$, respectively. The equal-altitude contour lines of zero and $-\pi$ magnitudes visually indicate how $\delta_{\rm magic}$ can be selected so that Eq. (1) is satisfied.

\begin{figure}
  \centering
  \includegraphics[width=3.25 in]{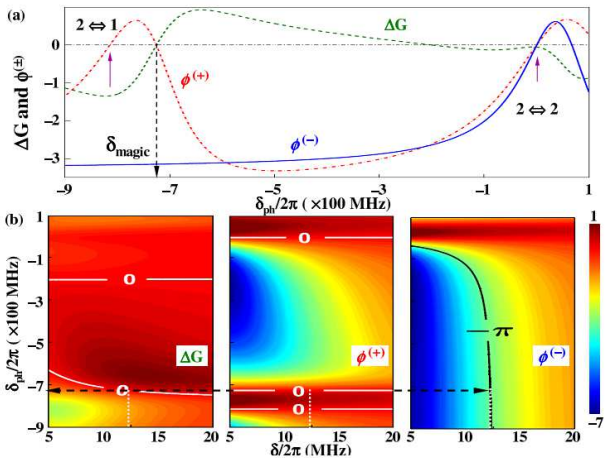}
  \caption{(a) Plots of $\Delta G=G^{(+)}-G^{(-)}$ and $\phi^{(\pm)}$ of the two circular components as functions of $\delta_{ph}$.  At $P_{ph}$ = 0.1 nW power a ``magic" $\delta_{ph}/2\pi=\delta_{\rm magic}/2\pi\approx-$720 MHz can be found, resulting in simultaneously $\Delta G=0$, $\phi^{(+)}=0$, and $\phi^{(-)}=-\pi$.
(b) Contour plots of $\Delta G$ and $\phi^{(\pm)}$.  Dashed arrows and dotted vertical lines jointly indicate the choices of $\delta_{ph}$ and $\delta$ where the three conditions in Eq. (1) are satisfied simultaneously. Parameters: $^{87}$Rb vapor ($\Gamma/2\pi$ = 6 MHz), density $n_0=1.4\times 10^9$/cm$^3$, $L$ = 1 cm, $I_P$ = 1.2 mW/cm$^2$, and $I_S$ = 0.05 nW/cm$^2$, respectively.}
\end{figure}

\vskip 10pt
\noindent The signal-field polarization rotation can be calculated by solving the Bloch-Maxwell's equation for each circular component, and then recombining them at the exit of the medium.  Figure 4 displays two contour plots that show the intensity of different polarization states of the signal field as a function of the phase-control light detuning and the normalized propagation distance in the medium.
The white dashed line indicates a ``magic" detuning $\delta_{\rm magic}$ at which the polarization state of the signal field changes from horizontal (${\bf H}$) at the entrance of the medium to vertical (${\bf V}$) at the exit of the medium.
Both contour plots are normalized with respect to the signal-field intensity at the entrance of the medium, i.e., $I_{\rm in}(z=0)$.  Correspondingly, we have $I_{\rm in}^{\rm (H)}$ = 1 (solid circle in left plot) and $I_{\rm in}^{\rm (V)}$ = 0 (solid circle in right plot) at the entrance $z/L=0$, but $I_{\rm out}^{\rm (H)}$ = 0 (open circle in the left plot) and $I_{\rm out}^{\rm (V)}\approx$ 3 (open circle in the right plot) at the exit $z/L$ =1.  The increase of the signal field at the exit is due to a small Raman gain.  Numerical calculations \cite{20} have shown that the PSKPS protocol described above, when combined with an electromagnetically induced transparency (EIT) \cite{20a} technique in which the signal field operates in an absorption mode, also exhibits similar CNOT/polarization gate-switching properties with less than 3-dB signal loss at the output.

\vskip 10pt
\noindent Experimentally, we use a warm $^{87}$Rb vapor to demonstrate the PSKPS-based polarization gate operation [see Fig. 2(b)].
The vapor cell has a length of 7.5 cm and a diameter of 2 cm.  It is shielded from ambient magnetic fields and is actively temperature stabilized to 322 K (density $n_0\approx 6\times 10^{11}$/cm$^2$). A solenoid is used to generate a weak magnetic field that provides an atomic quantization axis. We first optically pump the medium using a linearly polarized light field that couples the $|5S_{1/2}, F=2\rangle$ hyperfine manifold to the $|5P_{3/2}\rangle$ manifold [Fig. 2(a)]. Because the $|5P_{3/2}\rangle$ manifold radiatively decays to the $|5S_{1/2}, F=1\rangle$ manifold, in time there is a nearly complete population transfer to the $F=1$ ground-state manifold. A second optical pumping process may be carried out simultaneously using a separate $\sigma^{(+)}$ laser that couples the $|5S_{1/2}, F=1\rangle$ and $|5P_{1/2}, F^{'}=1\rangle$ manifolds, resulting in the medium being optically pumped to the $|5S_{1/2}, F = 1, m_{F}=+1\rangle$ state, which is the ideal starting state of the experiment. Due to Doppler broadening, there will be a small contribution from the $|5P_{1/2}, F^{'}=2, m_{F^{'}}=+2\rangle$ state that leads to a non-completely-closed optical-pumping cycle.  However, the contribution from this leakage transition can be substantially reduced by tuning the circularly polarized pumping field to the red side of the $|5P_{1/2}, F^{'}=1\rangle$ transition, which adds to the already large $|5P_{1/2}, F^{'}=1\rangle$ to $|5P_{1/2}, F^{'}=2\rangle$ hyperfine separation ( $\Delta_{hf}>$ 800 MHz).  Experimentally, however, we found that this secondary optical pumping process had a minimal effect on our results. Therefore, the experiments reported here only used the first optical-pumping process.
\begin{figure}
  \centering
  \includegraphics[width=3.25 in]{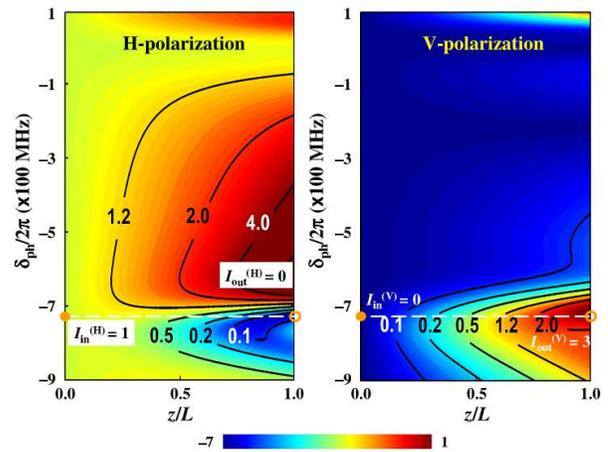}
  \caption{Equal-altitude-line-labeled color contour plots of the intensity of different signal-field polarization states as a function of $\delta_{ph}$ and $z/L$.  The white dashed line indicates the $\delta_{\rm magic}$ at which the polarization state of the signal field changes from ${\bf H}$ at the entrance ($z/L=0$, left panel) to ${\bf V}$ at the exit ($z/L=1$, right panel) of the medium. The one-photon pump detuning $\delta/2\pi\approx$12.5 MHz is given by the vertical dotted ling in Fig. 3 (other parameters are the same as in Fig. 3).}
\end{figure}
\vskip 10pt
\noindent Immediately following the optical pumping process, we turn on a strong, circularly polarized pump laser ($\ge$ 15 mW/cm$^2$ with beam diameter of 5 mm) that couples the $|5S_{1/2},F=1\rangle$ to  $|5P_{3/2},F^{''}=2\rangle$ transition with a one-photon detuning of $\delta/2\pi$ = 1.2 GHz above the $F^{''}=2$ level.  At the same time a linearly polarized signal field with a power of a few $\mu$W (beam diameter 1 mm) is switched on that couples the $|5P_{3/2},F^{''}=2\rangle$ to $|5S_{1/2},F=2\rangle$ transition with a two-photon detuning of about 700 kHz. These pump and signal fields are overlapped using a 50-50 beam splitter (BS) to form the two arms of a Mach-Zehnder interferometer. One arm will later be overlapped with a phase-control laser, and the other arm serves as a reference for phase analysis.  After exiting the cell, both arms of the Mach-Zehnder interferometer are joined together using a 50-50 BS. We set a small angle of less than 1$^{\rm o}$ between the pump and signal fields so that they are physically separated at a distance of about 1 m from the exit of the vapor cell. To introduce a Kerr phase shift we inject a weak circularly polarized phase-control field into one arm of the Mach-Zehnder interferometer using another 50-50 beam splitter. This gate-switching laser beam counter-propagates with respect to the signal light and has a power of 500 $\mu$W with a beam diameter of 5 mm (the signal beam diameter is 1 mm). The phase-control field couples the $|5S_{1/2},F=2\rangle\Leftrightarrow|5P_{1/2},F^{'}=1,2\rangle$ transition with $\delta_{ph}$ relative to the $5P_{1/2},F^{'}=2, m_F=+2$ level.

\begin{figure}
  \centering
  \includegraphics[width=3.25 in]{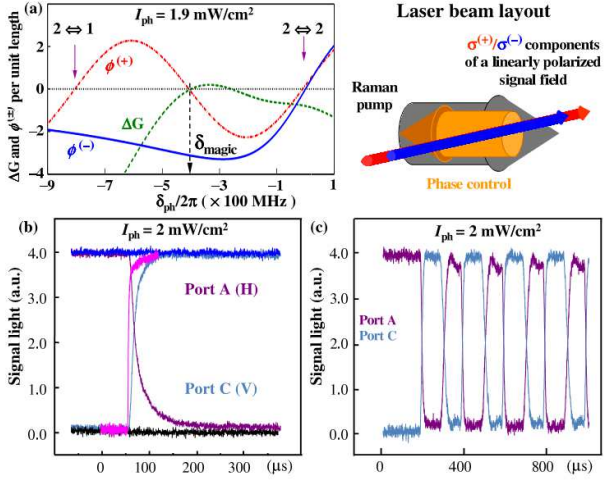}
  \caption{(a) Calculation of $\Delta G$ and $\phi^{(\pm)}$ for a Doppler-broadened medium used in experiment indicates that a ``magic detuning" can be found with a phase-control light intensity of $I_{ph}$ = 1.9 mW/cm$^2$.
Simulation parameters: $^{87}$Rb vapor ($\Gamma_D/2\pi$ = 320 MHz), $n_0=2.1\times 10^{11}$/cm$^3$, $\delta/2\pi=$ 1.2 GHz,
$L$ = 7.5 cm, and $I_P\approx$ 4 mW/cm$^2$. (b) Experimentally, high-fidelity orthogonal signal field polarization rotation occurs near $\delta_{\rm magic}/2\pi\approx-$400 MHz with $I_{ph}$ = 2 mW/cm$^2$ (0.5 mW power and 5 mm beam diameter). Also shown is fast polarization switching ($\sim$ 2 $\mu$s, the pink trace) with a slight timing adjustment of the pump and phase fields. (c) Repeated polarization switching by modulating the phase-control light. Upper-right: beam geometry.   }
\end{figure}
%
\vskip 5pt
\noindent In Fig. 5(a) we show numerical calculations of $\Delta G$ and $\phi^{(\pm)}$ of the two circular polarization modes using the parameters of the Doppler broadened $^{87}$Rb vapor used in our experiments. The phase-control light detuning is in reference to the $|2',+2\rangle$ state [see Fig. 2(a)]. At $\delta_{\rm ph}/2\pi=\delta_{\rm magic}/2\pi\approx-$ 400 MHz we find that $\Delta G\approx$ 0, $\phi^{(+)}\approx$ 0, and $\phi^{(-)}\approx\pi$ are simultaneously achieved with a polarization gate control field intensity $I_{ph}\approx$ 2 mW/cm$^2$ \cite{21}. At this detuning, complete signal-field polarization rotation and CNOT/polarization gate operation are expected and were observed experimentally.  Figure 5(b) shows that upon exit from the medium the polarization of the signal field is switched completely from ${\bf H}$ polarization to ${\bf V}$ polarization with nearly 100\% fidelity. The intensity of the polarization-switching field used in this high-fidelity CNOT/polarization gate operation is 2 mW/cm$^2$, which is equivalent to 0.5 nW, 3 $\mu$s optical pulse of wavelength $\lambda$ = 800 nm confined in a typical commercial photonic hollow-core fiber. In our experiment, the beam diameter of the signal light is 1 mm but the phase-control field diameter is 5 mm, indicating that the actual number of photons that contributed to the gate-swtichging operation is much smaller. We also note that slight adjustment in the timing of $\Omega_{P,S,ph}$ can substantially reduce the polarization switching time to about 2 $\mu$s [the pink trace in Fig. 5(b)], which is the limit of our system response time. In Fig. 5(c), we demonstrate rapid signal-light polarization switching by modulating the phase-control light.  This pulsed operation with high switching fidelity demontrates the viability of fast polarization switching operations using the PSKPS technique and the potential for advanced telecommunication applications.

\begin{figure}
  \centering
  \includegraphics[width=3.45 in]{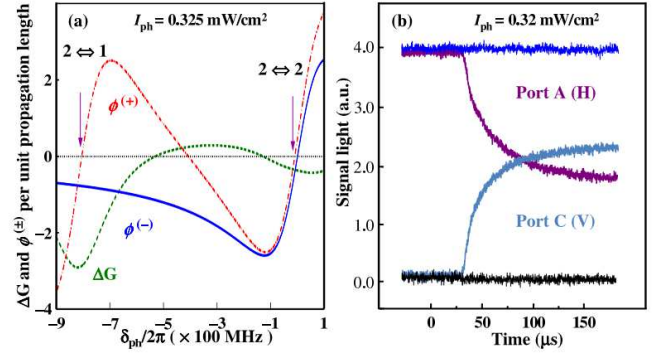}
  \caption{(a) Calculation of $\Delta G$ and $\phi^{(\pm)}$ for a Doppler-broadened medium indicates no ``magic detuning" can be found for phase-control light intensity of $I_{ph}=$ 0.325 mW/cm$^2$. The calculated polarization rotation is about 1 radian (near $\delta_{ph}/2\pi\approx$ - 450 MHz).
(b) Experimentally, signal field polarization rotation of 0.87 radian was observed at this intensity, agreeing with the numerical calculation shown in (a).}
\end{figure}

\vskip 10pt
\noindent One of the complications of room-temperature atomic vapors is the absence of a ``magic" detuning when the phase-control light intensity is less than some critical value (depending on other operational parameters such as the density, etc.). Due to the large Doppler broadening, contributions from nearby states in the Raman gain scheme, as well as in the EIT scheme, begin to adversely affect the polarization-rotation process. This can be seen from Fig. 6(a) where we numerically calculated $\Delta G$ and $\phi^{(\pm)}$ of the two circular polarization modes for $I_{ph}\simeq$ 0.32 mW/cm$^2$. Note that no $\delta_{\rm magic}$ can be found that can simultaneously satisfy Eq. (1). Experimentally, at this intensity the polarization of the signal light was still perfectly linear, but it was rotated only by about 0.87 radian [Fig. 6(b)], agreeing well with Fig. 6(a), but not enough to execute a complete polarization-gate operation. We point out that in the present experiment the phase-control light has a diameter of 5 mm. If an atom-confining photonic hollow-core fiber of 5 $\mu$m mode core were used instead, the power requirements for the phase-control light can be dramatically reduced. This demonstrates the superior technical performance and application potential of the PSKPS technique in the extreme-low-light-consumption telecommunications.
\vskip 10pt
\noindent While there are several avenues that can be followed to further improve the performance of the PSKPS protocol using atomic medium, from an application stand point it is desirable to use a solid state medium.  One of problems associated with solid media, however, is the fast relaxation rates of various states due to inhomogeneous broadening. This requires short pulse operation which adversely affects the effectiveness of the phase-control-light detuning.  However, the advantage in using a solid state medium is the much higher concentration of the four-level dopant and the possibility for highly confined guided-wave propagation in engineered microstructures with a few energy levels such as one-dimensional quantum well systems. Experimentally, optical confinement of weak phase shift light has already been tested using a photonic hollow core fiber filled with atomic vapor at about 80$^o$C \cite{22} using a ladder scheme. This parametric four-wave mixing scheme, however, has many complications and difficulties.  Aside from only producing very small phase shifts (several orders of magnitude smaller than required for a gate-switching operation), using a 5-ns short pulse with a ladder-scheme is prone to sporadic parametric photon emission. Such emission introduces significant frequency and directional instabilities which are extremely detrimental in very-low-light level optical communication applications.

\vskip 5pt
\noindent The high-fidelity polarization gate operation demonstrated here has the potential for applications in optical information communication and processing systems, and may lead to the development of critical building blocks for future all-optical extremely low power consumption next-generation telecommunication networks. Although the present experiment is still in the classical optical-field regime, the power consumption for the CNOT/polarization gate switching operations can be further reduced significantly by taking advantage of the strong optical confinement offered by photonic hollow fibers which significantly increases the effective phase-control field intensity.  Furthermore, combining the PSKPS technique demonstrated here with other schemes may lead to the ultimate fidelity-preserving CNOT/polarization gate operations critically important for future telecommunications and teleinformation processing. The most important future development, however, will be the demonstration of this PSKPS protocol using solid-state and engineered materials such as quantum wells and quantum dots arrays in the short-pulse regime.



\end{document}